\documentclass[aps,prd,twocolumn,10pt,groupedaddress]{revtex4-1}
\usepackage{amssymb}
\usepackage{graphicx}
\usepackage{amsmath}
\usepackage{hyperref}

\begin{document}

\title{Gravitational wave as probe of superfluid dark matter}
\author{Rong-Gen Cai}
\email{cairg@itp.ac.cn}
\author{Tong-Bo Liu}
\email{liutongbo@itp.ac.cn}
\author{Shao-Jiang Wang}
\email{schwang@itp.ac.cn}
\affiliation{CAS Key Laboratory of Theoretical Physics, Institute of Theoretical Physics, Chinese Academy of Sciences, Beijing 100190, China}
\affiliation{School of Physical Sciences, University of Chinese Academy of Sciences, No.19A Yuquan Road, Beijing 100049, P.R. China}
\date{\today}

\begin{abstract}
  In recent years, superfluid dark matter (SfDM) has become a competitive model of emergent modified Newtonian dynamics (MOND) scenario: MOND phenomenons  naturally emerge as a derived concept due to an extra force mediated between baryons by phonons as a result of axionlike particles condensed  as superfluid at galactic scales; Beyond galactic scales, these axionlike particles behave as normal fluid without phonon-mediated MOND-like force between baryons, therefore SfDM also maintains the usual success of $\Lambda$CDM at cosmological scales. In this paper, we use gravitational waves (GWs) to probe the relevant parameter space of SfDM. GWs through Bose-Einstein condensate (BEC) could propagate with a speed slightly deviation from the speed-of-light due to the change in the effective refractive index, which depends on the SfDM parameters and GW-source properties. We find that Five hundred meter Aperture Spherical Telescope (FAST), Square Kilometre Array (SKA) and International Pulsar Timing Array (IPTA) are the most promising means as GW probe of relevant parameter space of SfDM. Future space-based GW detectors are also capable of probing SfDM if a multimessenger approach is adopted.
\end{abstract}
\maketitle

\section{Introduction}

Despite the success of $\Lambda$CDM model at linear order and cosmological scales, there are two discordances that haunt the cosmologists and astronomers for decades: the galactic discordances and cosmic discordances. We will argue below that recently proposed superfluid dark matter (SfDM) scenario is capable of shedding some light on both galactic and cosmic discordances.

The galactic discordances lie in the semiempirical laws that govern galactic dynamics as results of either cold dark matter (CDM) or modified Newtonian dynamics (MOND) \cite{Milgrom:1983ca,Milgrom:1983pn,Milgrom:1983zz}. On the one hand, the very existence of DM is irrebuttable:
\begin{enumerate}
  \item At cosmic scales, the observations from the big bang nucleosynthesis (BBN), the cosmic microwave background (CMB) and the large scale structure (LSS) all point to some form of nonbaryonic gravitating mass, and this is the scale where MOND failed miserably.
  \item At cluster scales, the discovery of bullet cluster with offset mass distributions between the baryonic mass from optical and x-ray observations and the nonbaryonic mass from weak lensing provides almost direct proof \cite{Clowe:2006eq} of the very existence of DM.
  \item At galactic scales, the flatness of galaxy rotation curves clearly evince a mass discrepancy between baryonic matter and dynamical matter, of which the discrepancy is usually attributed  to the so-called DM.
\end{enumerate}

However, the direct N-body simulations with the use of CDM encounter with some small-scale anomalies (see recent review \cite{Tulin:2017ara} and references therein) like core-cusp, missing satellite and too-big-to-fail problems. Although all of these small-scale anomalies can be alleviated in the paradigm of Bose-Einstein condensate (BEC) DM \cite{Suarez:2013iw} with (known as self-interacting DM) or without self-interactions (known as fuzzy DM), on the other hand, the MOND still seems well established \cite{Famaey:2011kh} due to a critical acceleration scale written in the data that is otherwise unnatural to be seen in scale-free CDM :
\begin{enumerate}
  \item Globally, regardless of the specific distributions of baryonic mass along radial direction of galaxies, the asymptotic circular velocity is solely correlated with the total enclosed baryonic mass, which is known as the baryonic Tully-Fisher relation (BTFR) \cite{Tully:1977fu,McGaugh:2000sr} derived exactly from MOND.
  \item Locally, the observed distribution of baryonic mass predicts a radial acceleration that is strongly correlated with that traced by rotation curves, which is known as the mass discrepancy-acceleration relation (MDAR) \cite{McGaugh:2004aw,McGaugh:2014xfa,McGaugh:2016leg,Lelli:2017vgz} realized recently from MOND \cite{Milgrom:2016ogb} as well.
  \item Besides the BTFR and MDAR along with other Kepler-like galactic laws \cite{Famaey:2011kh} leading to the same critical acceleration scale, the galaxy rotation curve can be made universal \cite{Verheijen1999} if one properly normalizes the radial distance, regardless as to whether the galaxy is of high surface brightness (HSB) or low surface brightness (LSB). This universal rotation curve (URC) \cite{PersicSalucci} once again indicates that somehow the dynamics know intimately about the distribution of light, which will be too fine-tuning if DM is told to do the same thing.
\end{enumerate}

It seems like we are in a dilemma \cite{McGaugh:2014nsa} regarding to above galactic discordances, however, the SfDM provides us a hybrid way out by mimicking MOND phenomenons from axionlike particles condensed as superfluid at galactic scales, and at the same time maintaining the DM nature at cluster and cosmic scales. There are currently two kinds of models of SfDM that can produce MOND-like interaction with \cite{Berezhiani:2015bqa} or without \cite{Khoury:2016ehj} the help of the excited phonons from the condensed superfluid. See also \cite{Cai:2015rns} for an explanation of MOND critical acceleration scale by coupling  SfDM  to Dirac-Born-Infeld (DBI) dark energy.

Apart from the reconciliation of galactic discordances, SfDM might also be capable of alleviating the cosmic discordances. The cosmic discordances lie in the inconsistent measurements between CMB and LSS. On the one hand, the Hubble expansion rate inferred from CMB observation \cite{Ade:2015xua} is usually smaller than the local measurement from type-IA supernova \cite{Riess:2016jrr}; On the other hand, the matter fluctuations extrapolated from CMB constraints \cite{Ade:2015xua} into late-time regime is larger than that expected from low-redshift LSS observations \cite{Troxel:2017xyo,Abbott:2017wau}. Recently in \cite{Anand:2017wsj} the cosmic shear/bulk viscosity is shown clearly to be a natural and economic reconciliation of this CMB-LSS inconsistency between high redshift, large scale and low redshift, small scale. In SfDM scenario, the galaxy is within the superfluid phase with vanishing viscosity, however beyond galaxy cluster scales, those self-interacting axionlike particles are in normal fluid phase with viscosity that can be made in principle to reconcile the CMB-LSS tensions and even the cosmic acceleration \cite{Atreya:2017pny}. We will explore this possibility elsewhere in future work.

In this paper, we will adopt the recent proposal \cite{Dev:2016hxv} using the velocity change of GWs to probe the parameter space of BEC DM, which will be briefly reviewed in Sec. \ref{sec:GWprobe}. The difference here is that the BEC DM is SfDM with MOND emerging  at galactic scales. Both of SfDM models  considered in \cite{Berezhiani:2015bqa,Khoury:2016ehj} are estimated in Sec. \ref{sec:SfDM} with further considerations of two-fluid phases \cite{Hodson:2016rck} and baryon effect. In Sec. \ref{sec:observation}, the observational perspectives of different SfDM models are presented along with a discussion on Shapiro time delay between GWs and photons due to the effect of gravitational lensing. We summarize our result in Sec. \ref{sec:conclusion}.

\section{Gravitational-wave probe} \label{sec:GWprobe}

In \cite{Sabin:2014bua}, Sabin \textit{et al}. have shown that spacetime distortions can produce phonons in BEC. Thus, we can apply it to the detection of GWs. The effective metric of the excitations on the flat spacetime metric is given by
\begin{align}
g_{\mathrm{eff}}=\frac{\bar{n}^{2}}{\bar{c}_{s}(\bar{\rho}+\bar{P})}\mathrm{diag}(-\bar{c}_{s}^{2},\ 1,\ 1,\ 1),
\end{align}
where $\bar{n}$ is the mean number density of the background field, and the differential of mean pressure $\bar{P}$ with respect to the mean energy density $\rho$ gives rise to the square of the mean speed of sound, $\bar{c}_s^2=\mathrm{d}\bar{P}/\mathrm{d}\bar{\rho}$. So, the solutions of the Klein-Gordon equation with this effective metric describe massless excitations propagating with the speed of sound $\bar{c}_{s}$. As a result, we can obtain the following dispersion relation
\begin{align}
\omega_{k}=\bar{c}_{s}|\mathbf{k}|,
\end{align}
where $\omega_{k}$ is the frequency of the mode and the momentum of mode is denoted by $\mathbf{k}$.

In order to obtain the change in the speed of GWs, we must calculate the refractive index of GWs when scattering off scatters inside the medium. Here we follow \cite{Dev:2016hxv}, and apply the optical theorem, which links the index of refraction, $n_{g}$ to the forward scattering amplitude, $f(0)$ as
\begin{align}
n_{g}=1+\frac{2\pi \bar{n}f(0)}{k^{2}},
\end{align}
where $\bar{n}=\bar{\rho}/m$ is the mean number density of scatterers inside the medium and $k$ is the wave number of the incident wave. Since the exchange of energy comes along with the scattering of incident GWs, the forward scattering $\bar{n}f(0)$ is then expressed in terms of  the energy density of the GWs as well as that of the phononic excitations in the ground state. Therefore, the effective refractive index is given by
\begin{align}
n_{g}=1+\frac{\Delta k^{2}}{2\omega_\mathrm{GW}^{2}},
\end{align}
where, $\Delta k$ is the change of the wave number of the incident GWs and $\omega_\mathrm{GW}=2\pi f$ stands for the angular frequency. As noted in \cite{Dev:2016hxv}, this effect could be enhanced sizably due to the huge occupation number in the ground state and long-range correlations of the condensate. Therefore, it can be used for us to probe the SfDM with GWs. In the case of ordinary CDM, this effect is dramatically small and can be neglected. In the following context, we will explore the relation between the energy density of the GWs and that of the phononic excitations in the ground state as mentioned above.

To proceed, we consider the GWs produced at a distance $D$ from Earth, which is outside the galaxy. The typical energy density we adopt is
\begin{align}
\rho_\mathrm{GW}=\frac{1}{4}M_\mathrm{Pl}^{2}\omega_\mathrm{GW}^{2}h^{2},
\end{align}
where $h$ is the strain of GWs, and the Planck mass is related to Newton constant as $M_\mathrm{Pl}^{2}=1/8\pi G_\mathrm{N}$ by convention. The propagation of the GWs through the DM halo will result in the relative change in its wave number, which  can be calculated as
\begin{align}
\frac{\Delta\rho}{\rho_\mathrm{GW}}=2\frac{\Delta k}{\omega_\mathrm{GW}},
\end{align}
where $\Delta\rho$ represents the exchange of the energy density between GWs and phonons, and will be replaced with the energy density required for the phononic excitations, which we shall discuss later.

Next, we follow \cite{Sabin:2014bua} and assume that  the condensate is contained in a 1-dimensional cavity trap. The energy spectrum of the massless modes after imposing hard-wall boundary conditions, is then given by
\begin{align}
\omega_{l}=\frac{l\pi \bar{c}_{s}}{\langle D_{\mathrm{halo}}\rangle},
\end{align}
where $l\in \{1,2, \cdots\}$ and the denominator $\langle D_{\mathrm{halo}}\rangle=4R/\pi$ is the cavity length analog for GWs propagating through a spherical halo with radius $R$. In order to excite the massless modes inside the medium, the minimum energy density that we need is then given by the product of the number density of phonons and the energy difference between the closest modes
\begin{align}
\Delta \rho\equiv \bar{n}\Delta\omega=\frac{\bar{n}\pi^{2}\bar{c}_{s}}{4R},
\end{align}
where $\Delta\omega\equiv\omega_{l+1}-\omega_{l}$. This $\Delta \rho$ will compensate the change in energy density of the GWs encountering a DM halo as we noted above.

The average fraction of distance the GWs propagating through the halo with a reduced speed $c_{g}=1/n_{g}$ is given by
\begin{align}
x\equiv\frac{\langle D_{\mathrm{halo}}\rangle}{D}=\frac{4R}{\pi D}.
\end{align}
The effective speed of GWs can then be defined as
\begin{align}
c_{\text{eff}}\equiv\frac{D}{\Delta\tau}=\frac{c_{g}}{x+(1-x)c_{g}},
\end{align}
where $\Delta\tau=xD/c_{g}+(1-x)D$ represents the proper time that GWs take to propagate from the source location to the detector. Therefore, the change in the speed of the GWs due to the change of refractive index, compared with the speed of light in vacuum, which we adopt as $c=1$,  is given by
\begin{align}\label{eq:dcg}
\delta c_{g}\equiv 1-c_{\text{eff}}=\frac{x\delta n_{g}}{1+x\delta n_{g}},
\end{align}
where
\begin{align}\label{eq:dng}
\delta n_{g}\equiv n_{g}-1=\frac{\bar{n}^{2}\bar{c}_{s}^{2}}{128M_\mathrm{Pl}^{4}R^{2}f^{4}h^{4}},
\end{align}
is the change of refractive index due to the propagation of the GW through the BEC medium. The above expression depends not only on the parameters of the GW, like the frequency $f$ and characteristic strain $h$, but on the profile of SfDM as well, like the enclosed mass of the DM halo $M$, the mass of the axionlike particle $m$ and the characteristic energy scale $\Lambda$, which are encoded in the expression of $\bar{n}$ and $\bar{c}_{s}$. So we could apply \eqref{eq:dcg}\eqref{eq:dng} to constrain the ($m, \Lambda$) parameter space of SfDM for some fixed parameters of the target GWs detector, e.g.  $f$ and $h$. The results are given in the next section.

\section{Superfluid dark matter models} \label{sec:SfDM}

In this section, we will first study the SfDM model discussed in \cite{Berezhiani:2015bqa} in Sec. \ref{subsec:model1}, where the case of superfluid phase without baryons is studied in Sec. \ref{subsubsec:DMonly}, and the case of two-fluid phases without baryons is studied in Sec. \ref{subsubsec:twofluid}, and the case of superfluid phase including baryons is studied in Sec. \ref{subsubsec:addbaryon}. The second SfDM model \cite{Khoury:2016ehj} will be  studied in Sec. \ref{subsec:model2}, where only the case of superfluid phase without baryons is studied.

\subsection{Model A} \label{subsec:model1}

The general picture of model A \cite{Berezhiani:2015bqa} contains two parts: a MOND effective action describing SfDM phonons,
\begin{align}\label{eq:PX}
\mathcal{L}_\theta=P(X)=\frac23\Lambda(2m)^\frac32X\sqrt{|X|},
\end{align}
and a coupling term mediating MOND force between baryons,
\begin{align}\label{eq:XB}
\mathcal{L}_{\theta b}=-\frac{\alpha\Lambda}{M_\mathrm{Pl}}\theta\rho_b.
\end{align}
Here in \eqref{eq:PX}, $P(X)$ is the pressure from the effective-field-theory point-of-view in nonrelativistic regime at the lowest order in derivative, and $X\equiv\dot{\theta}-m\Phi-(\mathbf{\nabla}\theta)^2/2m$ describes superfluid phonons $\theta(t,r)=\mu t+\phi(t,r)$ expanded at constant chemical potential $\mu$ with phonon excitations $\phi(t,r)$ in external gravitational potential $\Phi$. The interaction term \eqref{eq:XB} is a minimal coupling between phonon $\theta$ and baryon density $\rho_b$ with coupling constant $\alpha$. The model parameters $m$ and $\Lambda$ are the mass of SfDM particle and the characteristic energy scale, respectively.

How could the Lagrangian $\mathcal{L}=\mathcal{L}_\theta+\mathcal{L}_{\theta b}$ reproduce MOND at galactic scales ? Consider a static spherically-symmetric approximation, $\theta(t,r)=\mu t+\phi(r)$, $X(r)=\hat{\mu}(r)-\phi'^2(r)/2m$, $\hat{\mu}(r)\equiv\mu-m\Phi(r)$, the equation-of-motion (EOM) of $\theta$ can be readily integrated as
\begin{align}\label{eq:EOM1}
\sqrt{2m|X(r)|}\phi'(r)=\frac{\alpha M_b(r)}{8\pi M_\mathrm{Pl}r^2}\equiv\kappa(r),
\end{align}
where $M_b(r)\equiv\int_0^r 4\pi r^2\mathrm{d}r\rho_b(r)$ is total enclosed baryon mass at radius $r$. It was shown in \cite{Berezhiani:2015bqa} that only the $X<0$ branch admits a MONDian regime where $\kappa(r)\gg\hat{\mu}(r)$ with solution $\phi'(r)=\sqrt{k(r)}$. To see this, in this MONDian regime, the acceleration experienced by baryons from the phonon-mediated force
\begin{align}
a_\phi(r)=\frac{\alpha\Lambda}{M_\mathrm{Pl}}\phi'(r)
=\sqrt{\frac{\alpha^3\Lambda^2}{M_\mathrm{Pl}}a_\mathrm{N}}
\end{align}
can be made to match the MONDian acceleration $a_\mathrm{MOND}=\sqrt{a_0a_\mathrm{N}}$ as long as $\alpha^3=a_0M_\mathrm{Pl}/\Lambda^2$, where $a_0=1.2\times10^{-10}\,\mathrm{m/s^2}$ is the critical acceleration scale in MOND scenario. Remarkably $\alpha$ is of order unity for $\Lambda\sim\mathrm{meV}$, which together with $m\sim\mathrm{eV}$ also gives rise to a DM halo with mass $M\sim10^{12}\,M_\odot$ of realistic size $R\sim10^2\,\mathrm{kpc}$ as shown in \cite{Berezhiani:2015bqa}.

As noted in \cite{Berezhiani:2015bqa}, the effective action of form $X^{3/2}$ is specifically chosen to reproduce the MOND law at galactic scales. Condensate of this form behaves like superfluid with equation-of-state $P\sim\rho^3$, which under viral expansion $P=\kappa_BT\rho+g_2(T)\rho^2+g_3(T)\rho^3+\cdots$ suggests that, the SfDM particles have negligible 2-body interactions and interact primarily through 3-body processes. This can be compared to the usual BEC DM with equation-of-state $P\sim\rho^2$, which is governed by the two-body interactions. More strange forms of equation-of-state have been studied before in condense matter physics, like the unitary Fermi gas with effective action of form $X^\frac52$. Therefore the nonanalytic form of effective action $X^\frac32$ of SfDM is not that strange from the effective-field-theory point of view. In fact, SfDM model can be constructed in \cite{Khoury:2016ehj} for arbitrary $n$ with effective action $X^n$, please see Sec. \ref{subsec:model2} for an introductory discussion.

More comments on the condensation of SfDM. First of all, there is no explicit self-interaction term in original paper \cite{Berezhiani:2015bqa} of SfDM model, and the total effective action consists of an nonanalytic kinetic term $X^\frac32$ and a coupling term $\theta\rho_b$ between phonons and baryons. Therefore, the phrase ``strong self-interaction'' is referred to the quantum effect of Bose-Einstein condensation of axionlike particles. Second, the self-interaction is not that strong, just enough for axionlike particles thermalized at galactic scales. As you can see from Eqs.(11)-(14) in the original paper \cite{Berezhiani:2015bqa}, the lower bound for the interaction cross section satisfies the current constraints on the cross section of self-interacting dark matter (SIDM). However, as pointed out in \cite{Berezhiani:2015bqa}, SfDM is considerably different than SIDM, therefore each constraint must be carefully revisited. Third, it is not the strong self-interaction but the phonon-mediated attraction force between baryons that is responsible for the MOND law at galactic scales. The phonon-baryon coupling term itself has already softly broken the global $U(1)$ symmetry explicitly only at the $1/M_\mathrm{Pl}$ level and is therefore technically natural. Finally, such phonon-baryon coupling term can arise from baryons coupling to the vortex sector of superfluid, which would give rise to a $\cos\theta\rho_b$ operator, thereby breaking the continuous shift symmetry down to a discrete subgroup. When expanded around the state at finite chemical potential $\phi=\theta-\mu t$, such operator would give the phonon-baryon coupling term to leading order, albeit with an oscillatory prefactor. As pointed out in \cite{Berezhiani:2015bqa}, such phonon-baryon coupling term is treated as an empirical term in the effective action necessary to obtain the MOND phenomenon.

In the following three subsections, we will derive the SfDM profile under three different circumstances, whose superfluid halo radius will be extracted to estimate the mass density, number density and sound speed in \eqref{eq:dng}.

\subsubsection{Superfluid phase without baryons} \label{subsubsec:DMonly}

First, the equation-of-state (EOS) for the superfluid phase,
\begin{align}\label{eq:EOS1}
P=\frac{\rho^3}{12\Lambda^2m^6},
\end{align}
can be easily obtained from
\begin{align}\label{eq:superfluid}
P(X)&=\frac23\Lambda(2m)^\frac32X|X|^\frac32;\\
n(X)&=P'(X)=\Lambda(2m)^\frac32|X|^\frac12;\\
\rho(X)&=mn(X)=m\Lambda(2m)^\frac32|X|^\frac12,
\end{align}
where the mass density $\rho=mn$ is in nonrelativistic case, and the number density is calculated under grand canonical ensemble $n=P'(\mu)=P'(|X|)$.

Second, the hydrostatic equilibrium equation
\begin{align}\label{eq:hydrostatic1}
P(r)=\int_r^\infty\rho\mathbf{\nabla}\Phi\cdot\mathrm{d}\mathbf{r},
\end{align}
along with Poisson equation
\begin{align}\label{eq:Poisson1}
\mathbf{\nabla}^2\Phi(r)=4\pi G_\mathrm{N}\rho(r)
\end{align}
gives rise to
\begin{align}\label{eq:hydroPoisson}
\frac{P'(r)}{\rho(r)}=-\frac{4\pi G_\mathrm{N}}{r^2}\int_0^r\mathrm{d}r r^2\rho(r).
\end{align}
After replacing $\rho=mn=mP'(\mu)=mP'(|X|)$ on the left-hand side (LHS) and $\rho=mn=m\Lambda(2m)^\frac32|X|^\frac12$ on the right-hand side (RHS), the equation above leads to following profile equation,
\begin{align}\label{eq:profile0}
\frac{1}{r^2}\frac{\mathrm{d}}{\mathrm{d}r}\left(r^2\frac{\mathrm{d}}{\mathrm{d}r}|X(r)|\right)=-4\pi G_\mathrm{N}m^2\Lambda(2m)^\frac32|X(r)|^\frac12,
\end{align}
which can be made dimensionless by normalizing \footnote{There is a typo in the eq.(37) of \cite{Berezhiani:2015bqa}.}
\begin{align}
r&=b\xi;\\
|X(r)|&=X_0\Xi(\xi),
\end{align}
namely,
\begin{align}
\frac{1}{\xi^2}\frac{\mathrm{d}}{\mathrm{d}\xi}\left(\xi^2\frac{\mathrm{d}\Xi}{\mathrm{d}\xi}\right)=-\frac{4\pi G_\mathrm{N}m^2\Lambda(2m)^\frac32b^2}{X_0^\frac12}\Xi^\frac12.
\end{align}
Choosing
\begin{align}
b^4=\frac{X_0}{128\pi^2G_\mathrm{N}^2\Lambda^2m^7}=\left(\frac{\rho_0}{32\pi G_\mathrm{N}\Lambda^2m^6}\right)^2,
\end{align}
one arrives at the Lane-Emden equation
\begin{align}\label{eq:profile1}
\frac{1}{\xi^2}\frac{\mathrm{d}}{\mathrm{d}\xi}\left(\xi^2\frac{\mathrm{d}\Xi}{\mathrm{d}\xi}\right)=-\Xi^\frac12.
\end{align}
The Lane-Emden equation can be solved numerically upon given boundary conditions $\Xi(0)=1$ and $\Xi'(0)=0$ \footnote{An analytic fit in eq.(42) of \cite{Berezhiani:2015bqa} for $X(r)$ is not necessary.}. Therefore one can find the value $\xi_1$ with vanishing profile $\Xi(\xi_1)=0$, and hence the size of the SfDM halo, which is determined as $R=b\xi_1$.

Third, instead of fixing $b$ with the central DM mass density $\rho_0$, we want to use the total enclosed mass $M(R)$ of SfDM halo. To do this, rewriting \eqref{eq:hydroPoisson} as
\begin{align}\label{eq:master}
r^2|X'(r)|=-G_\mathrm{N}mM(r),
\end{align}
and evaluating at DM halo radius $R$, one finds
\begin{align}\label{eq:b}
b=\left(\frac{M(R)}{128\pi^2G_\mathrm{N}\Lambda^2m^6\xi_1^2\Xi'(\xi_1)}\right)^\frac15.
\end{align}
In this subsection, we always fix halo mass at a fiducial value $M(R)=10^{12}\,M_\odot$ denoted simply as $M$.

Now, we are ready to evaluate the change of effective refractive index in \eqref{eq:dng}
\begin{align}
\delta n_g = &\,5.04\times10^{-29}\left(\frac{m}{\mathrm{eV}}\right)^\frac{44}{5}
\left(\frac{\Lambda}{\mathrm{meV}}\right)^\frac{18}{5}\nonumber\\
&\left(\frac{M}{10^{12}\,M_\odot}\right)^\frac65\left(\frac{f}{\mathrm{mHz}}\right)^{-4}\left(\frac{h}{10^{-21}}\right)^{-4},
\end{align}
from
\begin{align}
R(M,m,\Lambda)&=b(M,m,\Lambda)\xi_1;\\
\bar{\rho}(M,m,\Lambda)&=M\left/\frac43\pi R(M,m,\Lambda)^3\right.;\\
\bar{n}(M,m,\Lambda)&=\bar{\rho}(M,m,\Lambda)/m;\\
\bar{c}_s^2(M,m,\Lambda)&=\bar{\rho}(M,m,\Lambda)^2/4\Lambda^2m^6.
\end{align}
The velocity of the GWs changes correspondingly according to \eqref{eq:dcg}, which also depends on source distance $D$, frequency $f$, and strain $h$. Throughout the paper, we use following illustrative configurations \cite{Chen:2016isk} of different GW detectors to present different GW sources :
\begin{align}
\hbox{LIGO} &: D=400\,\mathrm{Mpc}, f=35\,\mathrm{Hz}, h=10^{-21};\label{eq:LIGO}\\
\hbox{ET}   &: D=10^3\,\mathrm{Mpc}, f=10\,\mathrm{Hz}, h=10^{-23};\label{eq:ET}\\
\hbox{LISA} &: D=10^3\,\mathrm{Mpc}, f=10^{-3}\,\mathrm{Hz}, h=10^{-21};\label{eq:LISA}\\
\hbox{BBO}  &: D=10^3\,\mathrm{Mpc}, f=10^{-1}\,\mathrm{Hz}, h=10^{-24};\label{eq:BBO}\\
\hbox{IPTA} &: D=10^3\,\mathrm{Mpc}, f=10^{-8.5}\,\mathrm{Hz}, h=10^{-17};\label{eq:IPTA}\\
\hbox{FAST} &: D=10^3\,\mathrm{Mpc}, f=10^{-9}\,\mathrm{Hz}, h=10^{-18};\label{eq:FAST}\\
\hbox{SKA}  &: D=10^3\,\mathrm{Mpc}, f=10^{-9.5}\,\mathrm{Hz}, h=10^{-19}\label{eq:SKA}.
\end{align}
The results are presented in the first line of Fig. \ref{fig:dcg}, which will be summarized along with other models in section Sec. \ref{sec:observation}.

\subsubsection{Two-fluid phases without baryons} \label{subsubsec:twofluid}

There is an unsatisfactory in the calculations presented in subsection \ref{subsubsec:DMonly}. At galactic scales the axionlike particles are condensed as superfluid, while beyond galactic scales, the axionlike particles behave like normal fluid. In the case of superfluid phase alone, the SfDM halo is enclosed at a radius $R$ where SfDM mass density vanishes. However, if we consider both superfluid and normal-fluid phases \cite{Hodson:2016rck}, the SfDM halo should be enclosed at a smaller radius $R_c$ with nonvanishing mass density, where mass densities and pressures of both phases are continuous at that radius,
\begin{align}
\rho_s(R_c)&=\rho_n(R_c);\label{eq:matching1}\\
P_s(R_c)&=P_n(R_c),\label{eq:matching2}
\end{align}
where index $s$ stands for the superfluid phase, and the normal fluid denoted by $n$, whose profile is chosen as isothermal profile for concreteness and simplicity,
\begin{align}
\rho_n(r)=\rho_c\left(\frac{R_c}{r}\right)^2,
\end{align}
other DM profile like NFW profile can also be used but with more free parameters encountered. The goal is to solve the matching equations \eqref{eq:match1} \eqref{eq:match2} for $R_c$ and $\rho_c$.

The first matching condition \eqref{eq:matching1} is just
\begin{align}\label{eq:match1}
\rho_c=m\Lambda(2m)^\frac32\sqrt{X_0(b)\Xi\left(\frac{R_c}{b}\right)},
\end{align}
and the second one can be qualified by equating the superfluid pressure with the hydrostatic equilibrium equation for the pressure of normal fluid,
\begin{align}
\frac{\rho_c^3}{12\Lambda^2m^6}=\int_{R_c}^\infty\rho_n(r)\frac{G_\mathrm{N}M(r)}{r^2}\mathrm{d}r,
\end{align}
where the total enclosed mass $M(r>R_c)$ is computed by
\begin{align}
M(r)=M_c+\int_{R_c}^r4\pi r^2\mathrm{d}r \rho_n(r),
\end{align}
with abbreviation $M_c\equiv M(R_c)$. Therefore the second matching condition \footnote{There are typos in eq.(17) of \cite{Hodson:2016rck}.} is
\begin{align}\label{eq:match2}
\frac{\rho_c^2}{12\Lambda^2m^6}=\frac{G_\mathrm{N}}{3}\left(\frac{M_c}{R_c}+2\pi\rho_cR_c^2\right),
\end{align}

To solve the matching equations \eqref{eq:match1} \eqref{eq:match2}, one still needs to specify the total enclosed mass $M_c$, which is determined similarly as \eqref{eq:master} by
\begin{align}
R_c^2|X'(R_c)|=-4\pi G_\mathrm{N}mM_c
\end{align}
or in dimensionless form,
\begin{align}
bX_0\xi_c^2\Xi'(\xi_c)=-4\pi G_\mathrm{N}mM_c.
\end{align}
Hence $M_c$ can also be expressed as a function of $R_c$ by
\begin{align}\label{eq:Mc}
M_c=\frac{bX_0}{4\pi G_\mathrm{N}m}\left(\frac{R_c}{b}\right)^2\left|\Xi'\left(\frac{R_c}{b}\right)\right|.
\end{align}
It is worth noting that, $b(M,m,\Lambda)$ is still computed according to \eqref{eq:b} as function of $M,m,\Lambda$, where $M$ should be chosen properly so that the total enclosed SfDM halo mass $M_c=10^{12}\,M_\odot$. In fact, after solving Eqs \eqref{eq:match1}, \eqref{eq:match2} and \eqref{eq:Mc} for $R_c(M,m,\Lambda)$ and $\rho_c(M,m,\Lambda)$ as functions of $M,m\ \rm and\ \Lambda$, one will find that $M_c/M=0.0284204$, hence $M$ will be fixed as $3.5186\times10^{12}\,M_\odot$ in this subsection. The mean mass density is thus straightforward obtained as
\begin{align}
\bar{\rho}(M,m,\Lambda)=\frac{M_c(R_c(M,m,\Lambda),M,m,\Lambda)}{\frac43\pi R_c^3(M,m,\Lambda)},
\end{align}
and the mean number density and sound velocity in \eqref{eq:dng} follow similarly as in the superfluid phase without baryons,
\begin{align}
\bar{n}(M,m,\Lambda)&=\bar{\rho}(M,m,\Lambda)/m;\\
\bar{c}_s^2(M,m,\Lambda)&=\bar{\rho}(M,m,\Lambda)^2/4\Lambda^2m^6.
\end{align}
The results for the change of GW velocity are presented in the second line of Fig. \ref{fig:dcg}, which will be summarized along with other models in Sec. \ref{sec:observation}.

\subsubsection{Superfluid phase including baryons} \label{subsubsec:addbaryon}

There is another unsatisfactory in the calculations presented in subsection \ref{subsubsec:DMonly}. The baryons also contribute the Poisson equation \eqref{eq:Poisson1},
\begin{align}\label{eq:Poisson2}
\mathbf{\nabla}^2\Phi(r)=4\pi G_\mathrm{N}(\rho_s(r)+\rho_b(r)),
\end{align}
thus influence the hydrostatic equilibrium equation \eqref{eq:hydrostatic1} or \eqref{eq:hydroPoisson} for the pressure of superfluid phase,
\begin{align}\label{eq:hydrostatic2}
\frac{P'_s(r)}{\rho_s(r)}=-\frac{4\pi G_\mathrm{N}}{r^2}\int_0^r\mathrm{d}r r^2(\rho_s(r)+\rho_b(r)),
\end{align}
therefore the SfDM profile equation \eqref{eq:profile0},
\begin{align}\label{eq:profilesb}
\frac{1}{r^2}\frac{\mathrm{d}}{\mathrm{d}r}\left(r^2\frac{\mathrm{d}}{\mathrm{d}r}|X(r)|\right)=-4\pi G_\mathrm{N}(\rho_s(r)+\rho_b(r)),
\end{align}
will be changed accordingly with addition of baryons, even we assume that baryons are subdominated in halo.

It seems that \eqref{eq:profilesb} is difficult to solve without prior knowledge of baryon distribution $\rho_b(r)$. Fortunately, the coupling term \eqref{eq:XB} that gives rise to MONDian solution $\phi'(r)=\sqrt{\kappa(r)}$ directly connects SfDM with baryons,
\begin{align}
r^2|X(r)|=\frac{\alpha}{16\pi mM_\mathrm{Pl}}M_b(r),
\end{align}
therefore the baryon distribution can be expressed as
\begin{align}\label{eq:profileb}
\frac{\mathrm{d}}{\mathrm{d}r}\left(r^2|X(r)|\right)=\frac{\alpha}{4mM_\mathrm{Pl}}\rho_b(r).
\end{align}
Combining \eqref{eq:profilesb}, \eqref{eq:profileb} with dimensionless normalizations $r=b\xi$ and $|X(r)|=X_0\Xi(\xi)$, one finally arrives at the profile equation with baryon correction,
\begin{align}\label{eq:profile2}
\frac{1}{\xi^2}\frac{\mathrm{d}}{\mathrm{d}\xi}\left(\xi^2\frac{\mathrm{d}\Xi}{\mathrm{d}\xi}\right)=-\Xi^\frac12-\frac{\rho_0^\frac12}{\alpha m\Lambda}\frac{1}{\xi^2}\frac{\mathrm{d}}{\mathrm{d}\xi}\left(\xi^2\Xi\right).
\end{align}
Here the normalization constant $b$ is given as before by
\begin{align}
b^4=\frac{X_0}{128\pi^2G_\mathrm{N}^2\Lambda^2m^7}=\left(\frac{\rho_0}{32\pi G_\mathrm{N}\Lambda^2m^6}\right)^2,
\end{align}
or in terms of total enclosed mass by
\begin{align}
b=\left(\frac{M}{128\pi^2G_\mathrm{N}\Lambda^2m^6\xi_1^2\Xi'(\xi_1)}\right)^\frac15.
\end{align}
Nevertheless, the value for $\xi_1$ and $b$ are different with those in Sec. \ref{subsubsec:DMonly} due to baryon correction, and they should be determined after solving \eqref{eq:profile2}.

\begin{figure}
  \includegraphics[width=8cm]{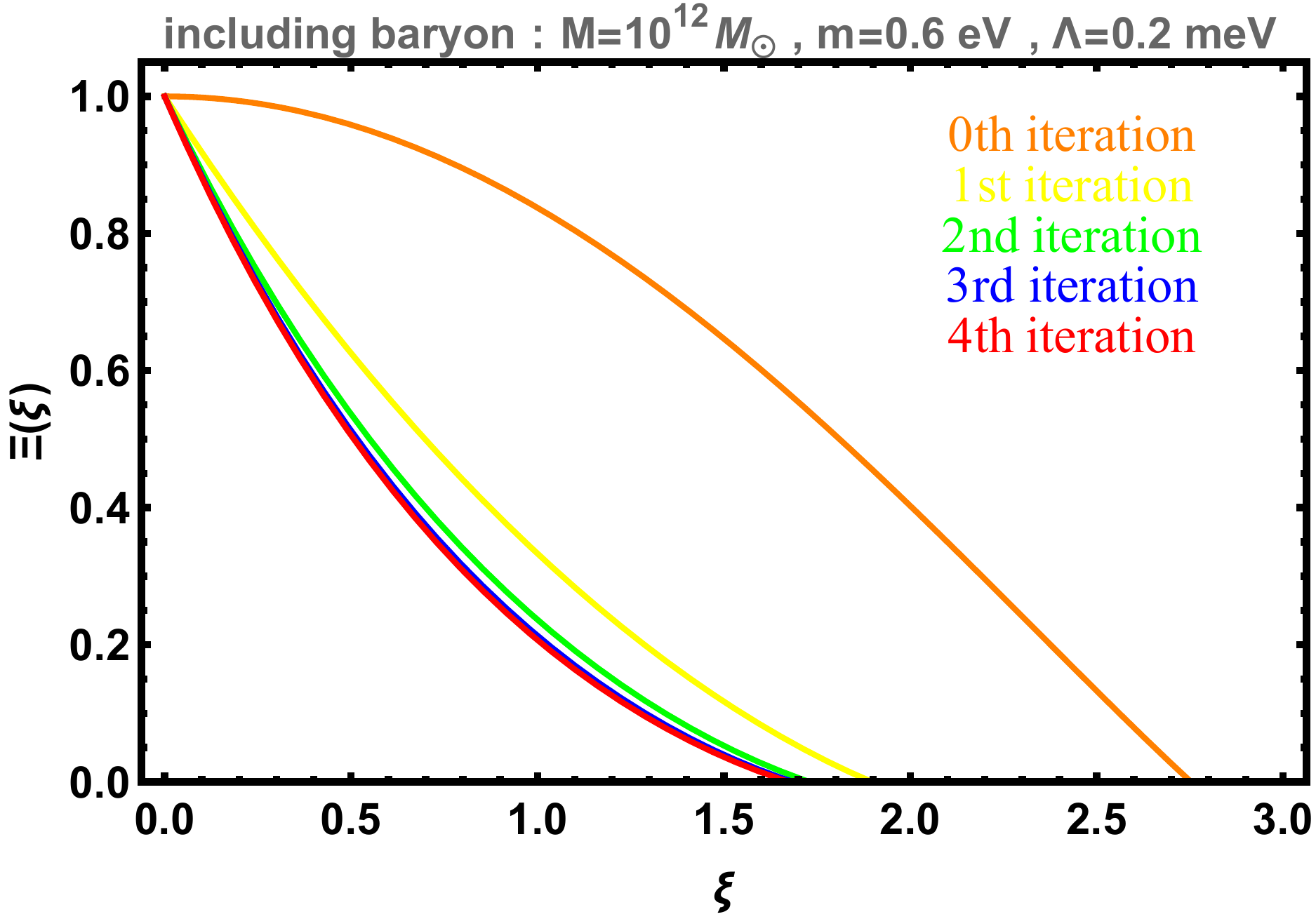}\\
  \caption{A demonstration of iteration algorithm for solving the profile equation in the superfluid phase of model A with addition of baryons. The parameters we choose are $M=10^{12}\,M_\odot, m=0.6\,\mathrm{eV}, \Lambda=0.2\,\mathrm{meV}$, respectively. Fourth iteration is enough for the profile of $\Xi(\xi)$ to stay fixed, as denoted by the red line.}\label{fig:iteration}
\end{figure}

Solving \eqref{eq:profile2} is new to our knowledge (see \cite{Berezhiani:2017tth} for more details on baryon-phonon coupling), and it is also tricky because $\rho_0$ cannot be specified for given $M, m, \Lambda$ without input value of $b$, which itself depends on the solution of \eqref{eq:profile2} through $\xi_1$. We propose here an iteration algorithm described below:
\begin{enumerate}
  \item solving \eqref{eq:profile2} without the baryon correction term, and obtaining the $0$th iteration solution $\Xi_{(0)}(\xi)$, then locating the values $\xi_1^{(0)}$ and $\Xi'_{(0)}(\xi_1^{(0)})$, and hence obtaining $b^{(0)}$ and $\rho_0^{(0)}$ as the functions of $M, m, \Lambda, \xi_1^{(0)}, \Xi'_{(0)}(\xi_1^{(0)})$;
  \item solving \eqref{eq:profile2} with the presence of baryon correction and input value of $\rho_0^{(0)}$, and obtaining the $1$-st iteration solution $\Xi_{(1)}(\xi)$, then locating the new values of $\xi_1^{(1)}$ and $\Xi'_{(1)}(\xi_1^{(1)})$, and hence obtaining $b^{(1)}$ and $\rho_0^{(1)}$ as the functions of $M, m, \Lambda, \xi_1^{(1)}, \Xi'_{(1)}(\xi_1^{(1)})$ for next iteration;
  \item repeating second step until $|\xi_1^{(n)}-\xi_1^{(n-1)}|$ smaller than given small number, and SfDM halo radius is then  $R^{(n)}=b^{(n)}(M, m, \Lambda, \xi_1^{(n)}, \Xi'_{(n)}(\xi_1^{(n)}))\xi_1^{(n)}$.
\end{enumerate}
We demonstrate this iteration algorithm in Fig.\ref{fig:iteration}
with $M=10^{12}\,M_\odot, m=0.6\,\mathrm{eV}, \Lambda=0.2\,\mathrm{meV}$. As we can see, after four iterations the profile of $\Xi(\xi)$ stays fixed, and hence the solution of \eqref{eq:profile2} is found. Hence, the value for $\xi_1$ and $b$ are obtained as the final iteration values $\xi_1^{(n)}$ and $b^{(n)}$. Therefore, the change of effective refractive index in \eqref{eq:dng} goes parallel as in the subsection \ref{subsubsec:DMonly} from
\begin{align}
R(M,m,\Lambda)&=b(M,m,\Lambda)\xi_1;\\
\bar{\rho}(M,m,\Lambda)&=M\left/\frac43\pi R(M,m,\Lambda)^3\right.;\\
\bar{n}(M,m,\Lambda)&=\bar{\rho}(M,m,\Lambda)/m;\\
\bar{c}_s^2(M,m,\Lambda)&=\bar{\rho}(M,m,\Lambda)^2/4\Lambda^2m^6.
\end{align}
The results of velocity change of GW through superfluid phase with baryons are presented in the third line of Fig.\ref{fig:dcg}, which will be summarized along with other models in Sec. \ref{sec:observation}.

\subsection{Model B} \label{subsec:model2}

Although the model A presented in \cite{Berezhiani:2015bqa} enjoys serval appealing features:
\begin{enumerate}
  \item DM and MOND share a common origin as different phases of axionlike particles;
  \item The MOND  emerges without the need of additional degrees of freedom;
  \item The phonon of BEC DM is fully appreciated for generating MOND law among baryons;
  \item The idea of SfDM naturally distinguishes between galaxies and galaxy clusters,
\end{enumerate}
there are also some drawbacks,
\begin{enumerate}
  \item The finite temperature corrections are required to cure the instability of the wrong-sign kinetic term from perturbation around the MONDian solution at zero-temperature;
  \item The value of $a_0$, $\alpha$ and $\Lambda$ depends on temperature in such a way that their values at cosmic scales are four orders of magnitude deviated from those at galactic scales;
  \item The form of kinetic term of superfluid is of nonanalytic nature.
\end{enumerate}
Those motivate to propose another model for SfDM  \cite{Khoury:2016ehj}, which makes several differences as follows:
\begin{enumerate}
  \item The phonon excitations are no longer responsible for MOND, and thus the EOS of BEC can be of general type;
  \item The zero-temperature action is fully analytic in all field variables, and stable all by itself without finite temperature corrections;
  \item The MOND is realized universally for both DM and baryons.
\end{enumerate}

In \cite{Khoury:2016ehj}, Khoury proposed a next-to-leading order (NLO) term containing higher-derivative operators in addition to the leading order (LO) superfluid action plus minimal coupling term,
\begin{align}
\mathcal{L}_\mathrm{tot}&=\mathcal{L}_\mathrm{LO}+\mathcal{L}_\mathrm{NLO}+(-\Phi\rho_b);\\
\mathcal{L}_\mathrm{LO}&=\frac{\Lambda^4}{n}\left(\frac{X}{m}\right)^n;\\
\mathcal{L}_\mathrm{NLO}&=-\frac12Z^2(\partial\chi)^2
-M_\mathrm{Pl}^2(\mathbf{\nabla}\Phi)^2\left(\frac{1}{1+\chi^2}+\frac{(\mathbf{\nabla}X)^2}{9m^2a_0^2}\chi^2\right),
\end{align}
where $n=2$ for concreteness as standard BEC, and $X=\mu-m\Phi$ for the absence of phonon excitations and at finite chemical potential. The LO term expanded with $X=\mu-m\Phi$ at leading order, together with the minimal coupling term, gives rise to a contribution of form $-(\rho_s+\rho_b)\Phi$, where $\rho_s=\Lambda^4\left(\frac{\mu}{m}\right)^{n-1}$. Furthermore, the  ``symmetron'' field $\chi$ lives in an effective potential
\begin{align}
V(\chi)=M_\mathrm{Pl}^2(\mathbf{\nabla}\Phi)^2\left(\frac{1}{1+\chi^2}+\frac{(\mathbf{\nabla}\Phi)^2}{9a_0^2}\chi^2\right)
\end{align}
with $\mathbb{Z}_2$ symmetry $\chi\rightarrow-\chi$ spontaneously broken when the effective mass square
\begin{align}
m_\chi^2=2M_\mathrm{Pl}^2(\mathbf{\nabla}\Phi)^2\left(-1+\frac{(\mathbf{\nabla}\Phi)^2}{9a_0^2}\right)
\end{align}
flips a sign in the MONDian regime $|\mathbf{\nabla}\Phi|<3a_0$ with vacuum expectation value (VEV) as
\begin{align}\label{eq:VEV}
\chi=\pm\sqrt{\frac{3a_0}{|\mathbf{\nabla}\Phi|}-1}.
\end{align}
Expanding the NLO term around above VEV admits
\begin{align}
\mathcal{L}_\mathrm{NLO}\simeq-\frac{2M_\mathrm{Pl}^2}{3a_0}\left((\mathbf{\nabla}\Phi)^2\right)^\frac32
+\frac{M_\mathrm{Pl}^2}{9}\frac{(\mathbf{\nabla}\Phi)^4}{a_0^2},
\end{align}
where the second term is subdominated in the deep MOND regime $|\mathbf{\nabla}\Phi|\ll a_0$. Therefore in the deep MONDian and symmetry-breaking phase with VEV \eqref{eq:VEV}, the effective action to LO in gradients is
\begin{align}
\mathcal{L}_\mathrm{MOND}\simeq-\frac{2M_\mathrm{Pl}^2}{3a_0}\left((\mathbf{\nabla}\Phi)^2\right)^\frac32-(\rho_s+\rho_b)\Phi,
\end{align}
of which the EOM is of MONDian form,
\begin{align}
\mathbf{\nabla}\cdot\left(\frac{|\mathbf{\nabla}\Phi|}{a_0}\mathbf{\nabla}\Phi\right)=4\pi G_\mathrm{N}(\rho_s+\rho_b).
\end{align}
As one can see that, the new model of SfDM is different from the one presented in \ref{subsec:model1}. In \ref{subsec:model1}, only baryons experience extra phonon-mediated MONDian force that is larger than Newtonian force at galactic scales. However, in this subsection, both baryons and DM particles feel MONDian force steamed from the NLO term in symmetry broken phase within deep MONDian regime.

The SfDM profile was derived in \cite{Khoury:2016ehj} for the static, spherically-symmetric halo without baryons. The hydrostatic equilibrium equation for the pressure of SfDM is of MONDian form,
\begin{align}
\frac{X'(r)}{m}\equiv\frac{P'_s(r)}{\rho_s(r)}=-\sqrt{\frac{4\pi G_\mathrm{N}a_0}{r^2}\int_0^r\mathrm{d}r r^2(\rho_s(r)+\rho_b(r))},
\end{align}
since both DM and baryon are coupled to MOND gravity. However, unlike the model A in Sec. \ref{subsubsec:addbaryon} with a direct connection \eqref{eq:profileb} due to the phonon-mediated MOND force among baryons, the baryon contributions have to be ignored since there is no prior knowledge of baryon distribution $\rho_b(r)$, and one can only work out the SfDM profile in DM-only calculations \footnote{There is a typo in eq.(51) of \cite{Khoury:2016ehj}.}. Parallel to the calculations in Sec. \ref{subsubsec:DMonly}, the normalized variables $\Xi=X/X_0$ and $\xi=br$ are defined in such  a  way with
\begin{align}
b=(4\pi G_\mathrm{N}a_0\Lambda^4)^\frac13\left(\frac{X_0}{m}\right)^\frac{n-3}{3}
\end{align}
that the SfDM profile equation is dimensionless,
\begin{align}
\frac{1}{\xi^2}\frac{\mathrm{d}}{\mathrm{d}\xi}\left(\xi^2\left(\frac{\mathrm{d}\Xi}{\mathrm{d}\xi}\right)^2\right)=\Xi^{n-1},
\end{align}
which can be readily solved numerically upon given boundary conditions $\Xi(0)=1$ and $\Xi'(0)=0$. The $X_0$ in $b$ can be similarly expressed in terms of the enclosed SfDM halo mass as in Sec. \ref{subsubsec:DMonly} by noting that
\begin{align*}
M(r)=\int_0^r4\pi r^2\mathrm{d}r\rho_s(r)=\int_0^r4\pi r^2\mathrm{d}r \frac{(r^2X'^2)'}{4\pi G_\mathrm{N}a_0m^2r^2},
\end{align*}
which gives rise to
\begin{align}
X_0=m\left(\frac{a_0G_\mathrm{N}M}{\xi_1^2\Xi'(\xi_1)^2}\right)^\frac12
\end{align}
when evaluated at halo radius $R=b\xi_1$ with $\Xi(\xi_1)=0$.

Once we have the halo radius
\begin{align}
R=(4\pi G_\mathrm{N}a_0\Lambda^4)^\frac13\left(\frac{a_0G_\mathrm{N}M}{\xi_1^2\Xi'(\xi_1)^2}\right)^\frac{n-3}{6}\xi_1,
\end{align}
the calculations of velocity change of GWs go parallel as those in Sec. \ref{subsubsec:DMonly},
\begin{align}
R(M,\Lambda)&=b(M,\Lambda)\xi_1;\\
\bar{\rho}(M,\Lambda)&=M\left/\frac43\pi R(M,\Lambda)^3\right.;\\
\bar{n}(M,m,\Lambda)&=\bar{\rho}(M,\Lambda)/m;\\
\bar{c}_s^2(M,m,\Lambda)&=\frac{1}{n-1}\left(\frac{\bar{\rho}(M,\Lambda)}{\Lambda^4}\right)^\frac{1}{n-1}.
\end{align}
and the change of effective refractive index in \eqref{eq:dcg} is then given by
\begin{align}
\delta n_g = &\,8.673\times10^{-31}\left(\frac{m}{\mathrm{eV}}\right)^{-2}
\left(\frac{\Lambda}{\mathrm{eV}}\right)^\frac{32}{3}\nonumber\\
&\left(\frac{M}{10^{12}\,M_\odot}\right)^\frac76\left(\frac{f}{\mathrm{mHz}}\right)^{-4}\left(\frac{h}{10^{-21}}\right)^{-4}
\end{align}
The parameter space model B is estimated in \cite{Khoury:2016ehj} as $m\lesssim2\,\mathrm{eV}$ and $\Lambda\gtrsim2\times10^\frac{n-2}{2}\,\mathrm{eV}$.  The results for $\delta c_g$ with $n=2$ are presented in the last line of Fig.\ref{fig:dcg}, which will be summarized along with other models in section Sec. \ref{sec:observation}.

\section{Observational perspectives} \label{sec:observation}

In this section, we summarize the observational perspective of constraining different SfDM models presented in the last section.

First, how well have we know for the possible deviation of GW velocity from the speed-of-light ? Reference \cite{Moore:2001bv} put a bound $\delta c_g<2\times10^{-19}\sim2\times10^{-15}$ obtained from the absence of gravitational Cherenkov radiation for the observation of the highest energy cosmic rays. However, the direct observations of GW can also put bound on the GW velocity from three different ways:
\begin{enumerate}
  \item The simplest approach is to measure the arrival time of the GW from compact binary system and the electromagnetic (EM) waves from EM counterparts of that compact binary system, if we understand well enough about the intrinsic time-lag between GW emission and photons emission, which is often taken to be zero as \textit{ad hoc} estimation for the GW velocity, namely
      \begin{align}
      \frac{\delta c_g}{c}=\frac{c\delta t}{c\delta t+D}\simeq\frac{c\delta t}{D}
      \end{align}
      Here $\delta t$ is the arrival time difference between GW and gamma-ray burst (GRB), and $D=d_L/(1+z)^2$ is the physical distance estimated from the luminosity distance $d_L$ and redshift $z$, where $d_L$ can be directly obtained from standard siren and $z$ is obtained from multimessenger observations. Although lacking unambiguous evidences for the correlation between the Fermi-GRB event \cite{Connaughton:2016umz} and the GW150914 \cite{Abbott:2016blz} event, such distant GW events can in principle constrain the change of GW velocity down to $\delta c_g\leq10^{-40}\sim10^{-17}$ level \cite{Li:2016iww,Ellis:2016rrr,Branchina:2016gad}. Recent observed time delay $(+1.74\pm0.05)$ between the GW 170817 event \cite{TheLIGOScientific:2017qsa} and GRB 170817 event \cite{Goldstein:2017mmi,Savchenko:2017ffs} has put a stringent bound $-3\times10^{-15}<c_g<+7\times10^{-16}$ on the GW velocity \cite{Monitor:2017mdv}.
  \item Without the EM counterparts for GW events, the phase changes of GW waveform alone \cite{Will:1997bb,DelPozzo:2011pg} can bound the Compton wavelength of graviton from massive gravity (MG), namely
      \begin{align}
      h_\mathrm{MG}(f)&=Af^{-\frac76}\exp(i\Psi_\mathrm{MG}(f));\\
      \Psi_\mathrm{MG}(f)&=\Psi_\mathrm{GR}(f)-\frac{\pi^2DM}{\lambda_g^2(1+z)}(\pi Mf)^{-1},
      \end{align}
       or equivalently the graviton mass \cite{TheLIGOScientific:2016src} through the definition of Compton wavelength $\lambda_g=h/(m_gc)$. Matched filtering of the GW waveforms from inspiralling compact binaries can in principle constrain a frequency-dependent GW velocity, which manifests an offset in the relative arrival times at a detector, since the GW emitted at low frequency early during inspiralling stage will travel slightly slower than those emitted at high frequency later. However, transforming the constraint on the graviton mass to the constraint on the GW velocity is nontrivial due to the modified dispersion relation $\omega(k)$ from massive gravity that satisfies
       \begin{align}
       m_g^2c^4&=\hbar^2\omega(k)^2-\hbar^2k^2c^2;\\
       \frac{c_g^2}{c^2}&=\frac{k^2c^2}{\omega(k)^2},
       \end{align}
       or more complicated forms in other modified gravity \cite{Abbott:2017vtc}.
       Nevertheless, this way of constraining the GW velocity can never reach the precision that can be achieved easily from the joined measurements with EM counterparts.
  \item The joined measurement with EW counterparts of GW events can only make a radical estimation on the GW velocity, whose improvement relies on the well understanding of GRB emission relative to GW emission. There are other ways that do not heavily rely on the multimessenger observations and set an absolute upper limit on GW propagation speed. For example, \cite{Cornish:2017jml} gave a very loose bound $0.55c<c_g<1.42c$ with $90\%$ confidence on the GW velocity, if one notices that the GWs are arriving at the two detectors of LIGO with a time difference \cite{Blas:2016qmn}. However, with more GW events and more large worldwide network of detectors, the bound can be improved significantly. \cite{Cornish:2017jml} hence forecasts that just five GWs events by the LIGO-Virgo-Kagra network will constrain the GW velocity within $1\%$ precision. A second example is the strongly lensed GW events \cite{Collett:2016dey} that can be used to produce robust constraints on GW velocity at the $10^{-7}$ level without assuming vanishing emission lag in the source and without knowledge of the sky position of the inspiral event. Another example is to fully appreciate the longtime observations of eclipsing binaries with periodic signals under the so called \textit{phase lag test with eclipsing binaries} \cite{Bettoni:2016mij}, where the phase lag between the GW and the EM signals accumulates such amount that the dwarf binary system WDS J0651+2844 can be used to constrain the GW velocity at the level of few parts in a trillion.
\end{enumerate}

\begin{figure*}
  \centering
  \includegraphics[width=6.5cm]{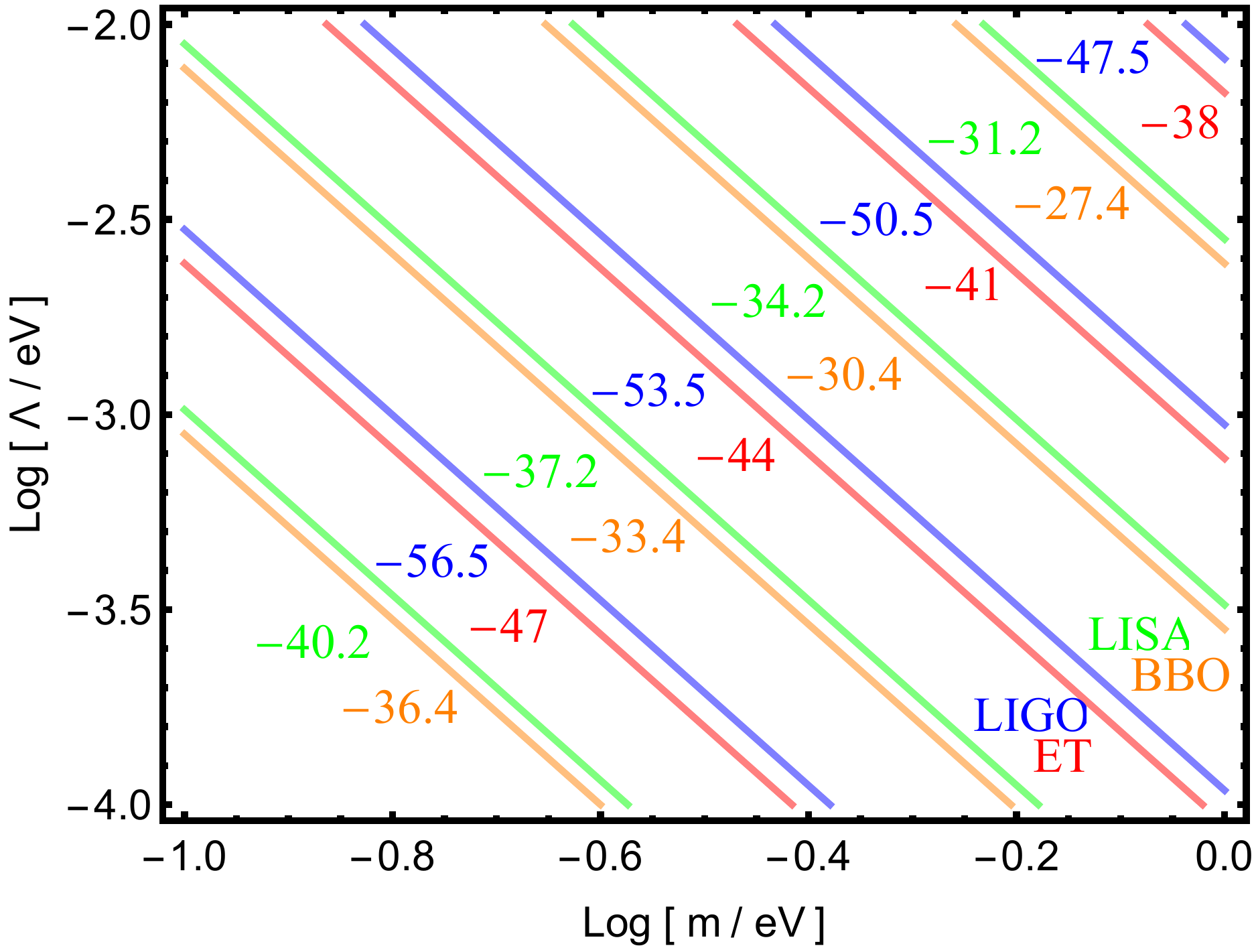}\qquad\qquad
  \includegraphics[width=6.5cm]{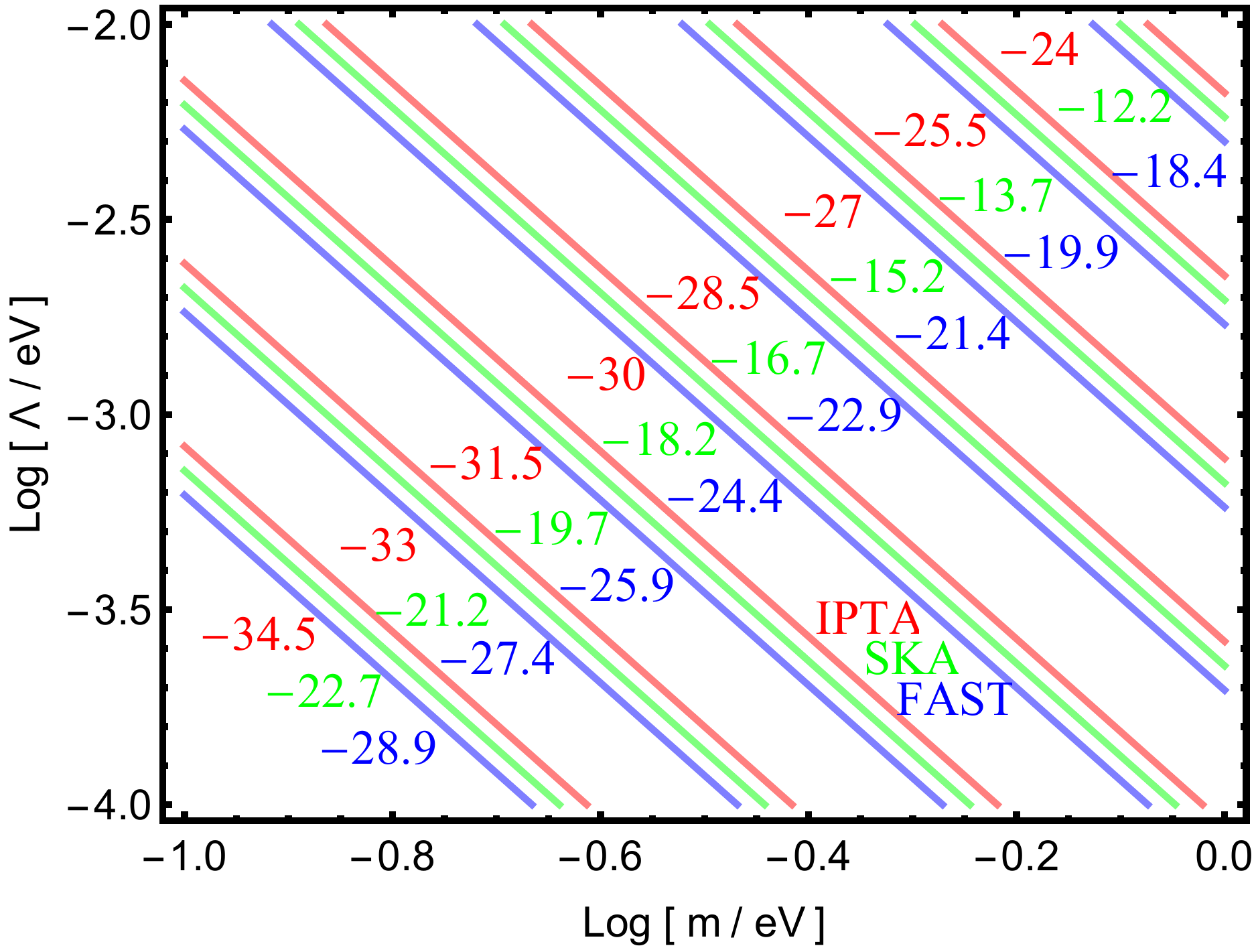}\\
  \includegraphics[width=6.5cm]{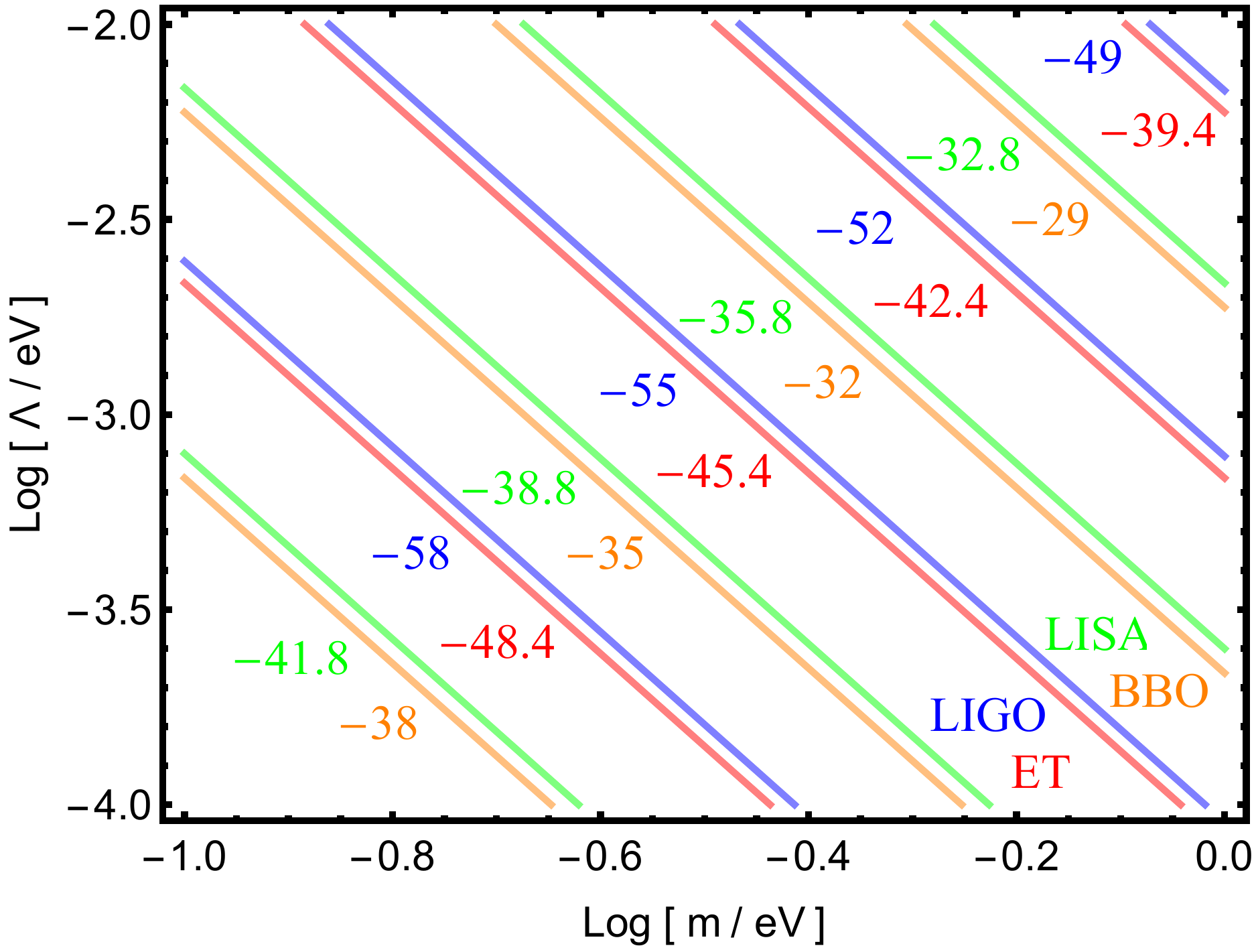}\qquad\qquad
  \includegraphics[width=6.5cm]{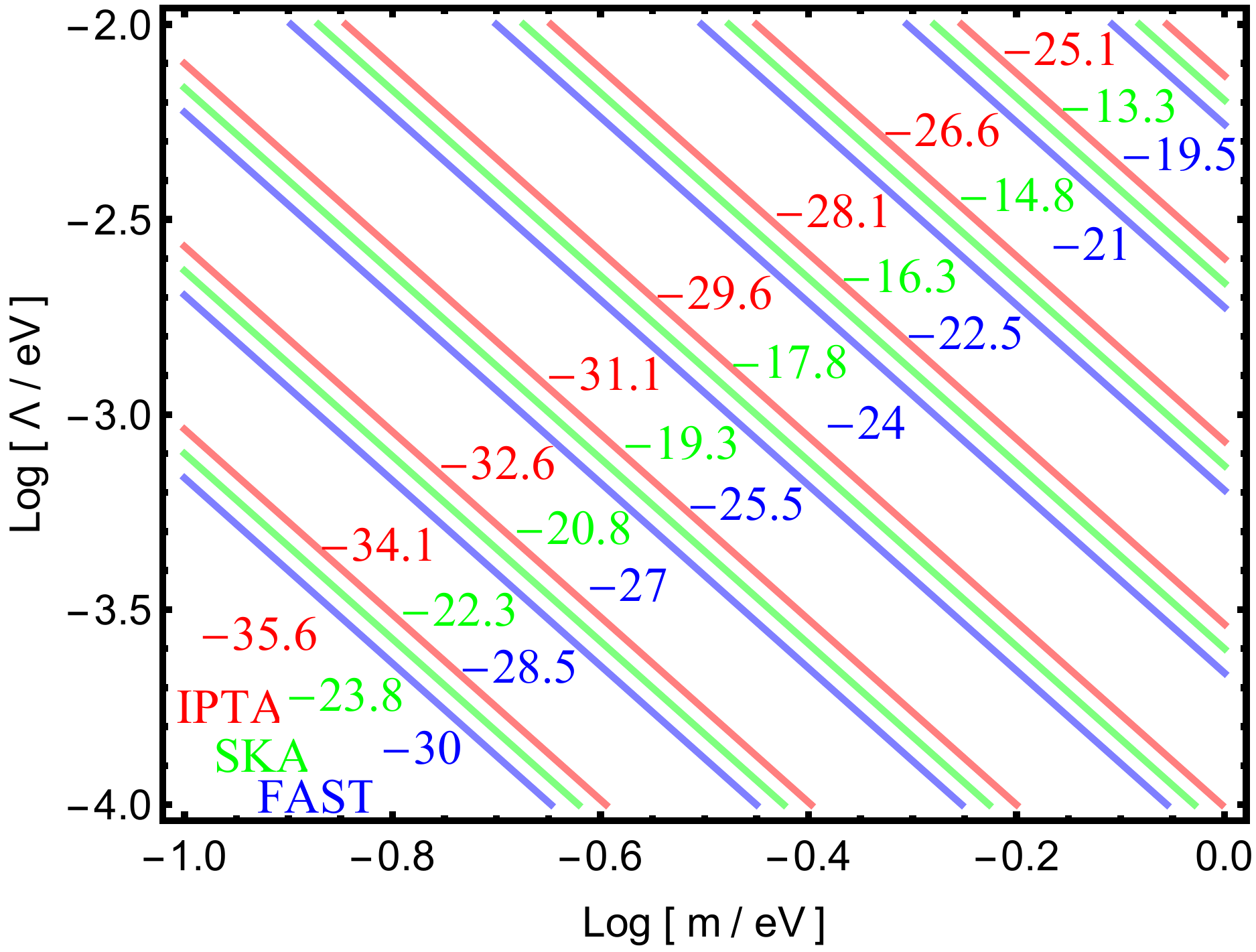}\\
  \includegraphics[width=6.5cm]{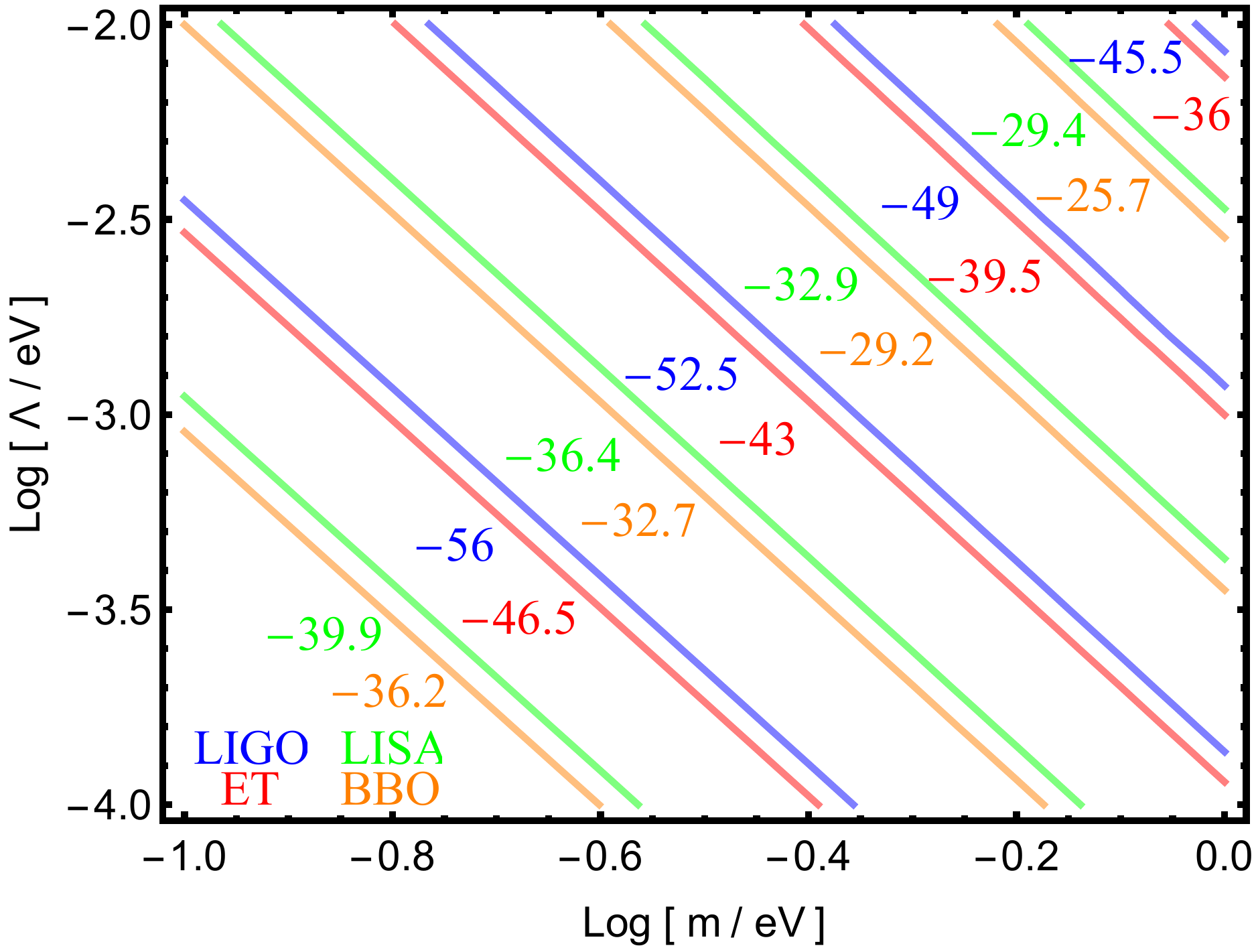}\qquad\qquad
  \includegraphics[width=6.5cm]{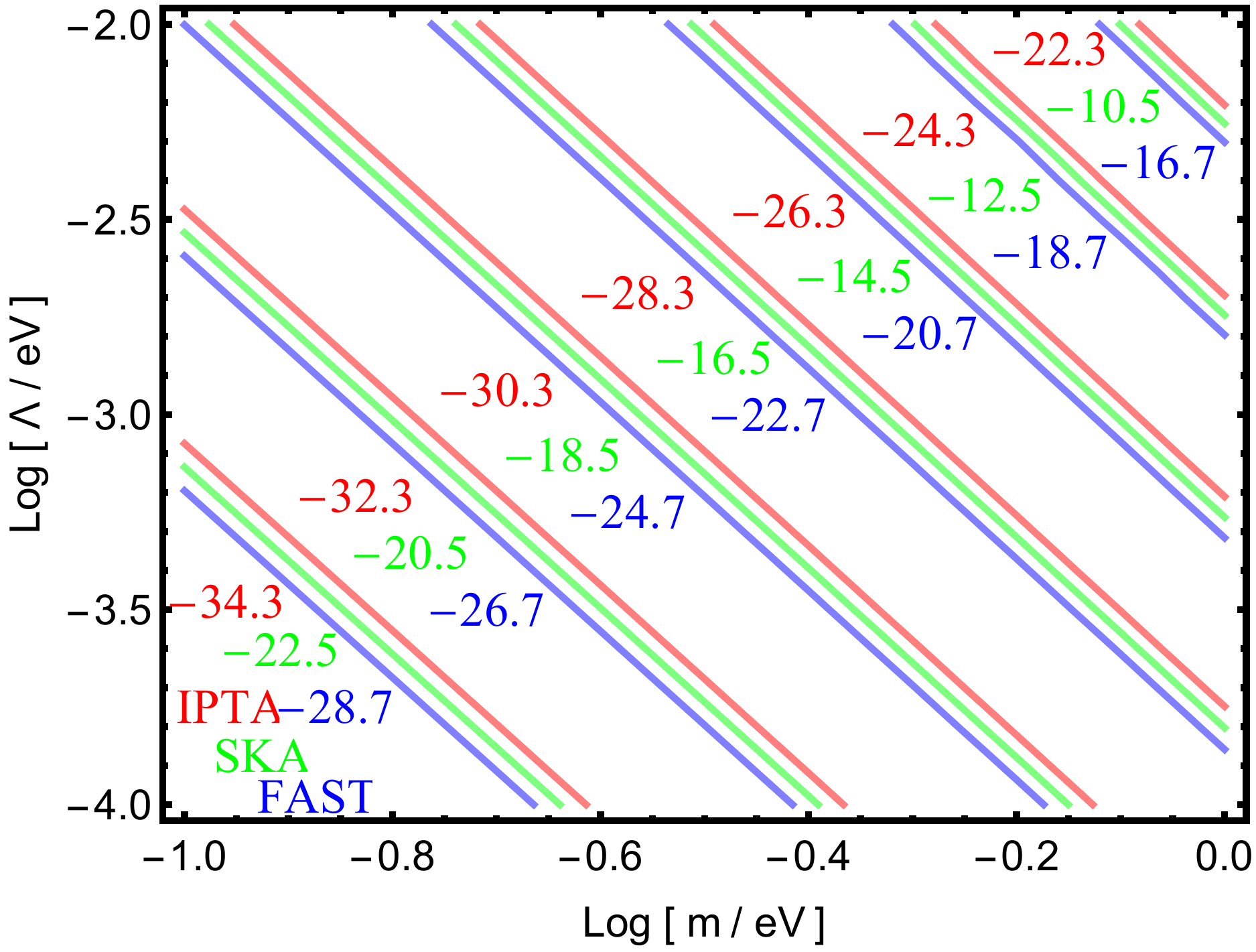}\\
  \includegraphics[width=6.5cm]{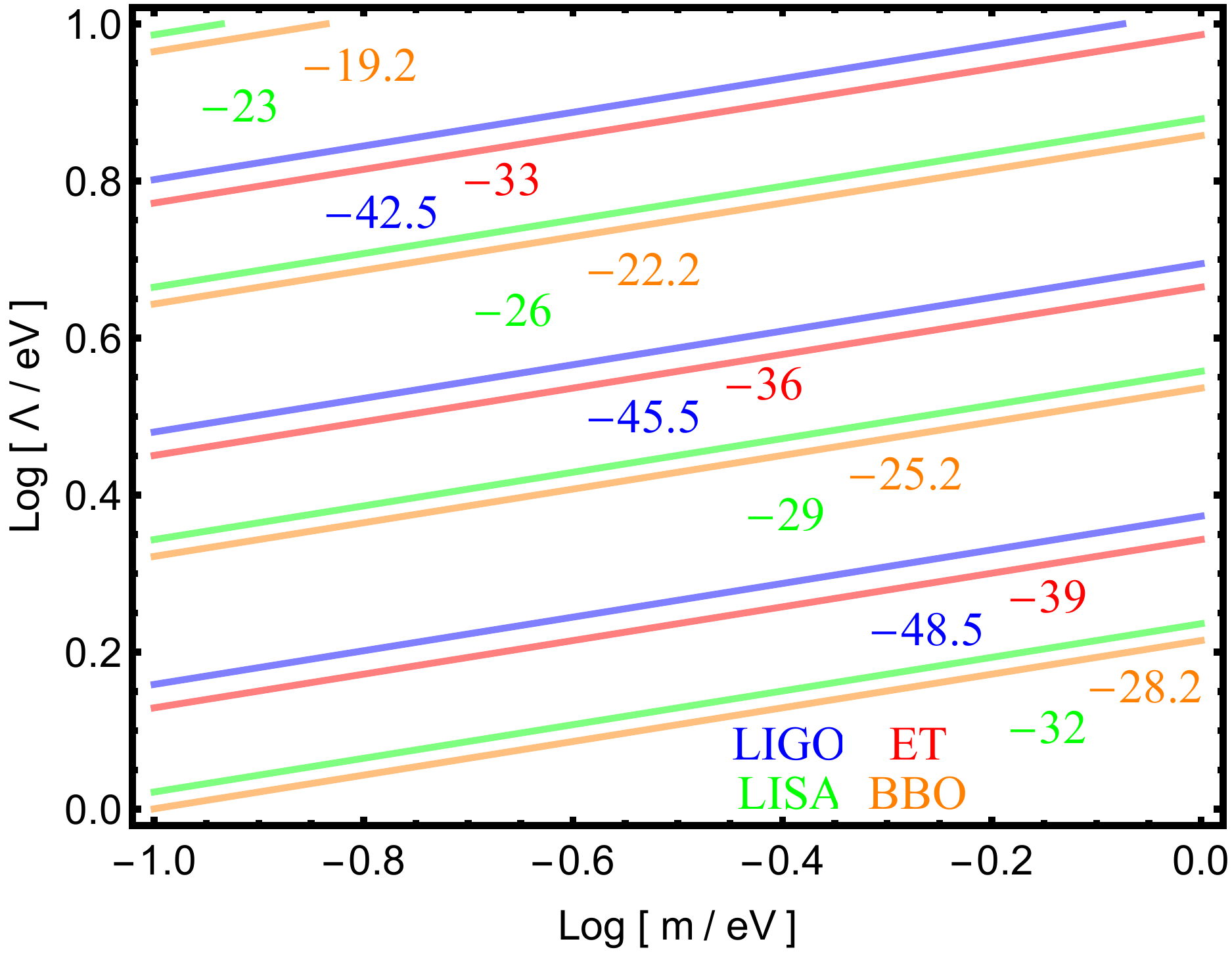}\qquad\qquad
  \includegraphics[width=6.5cm]{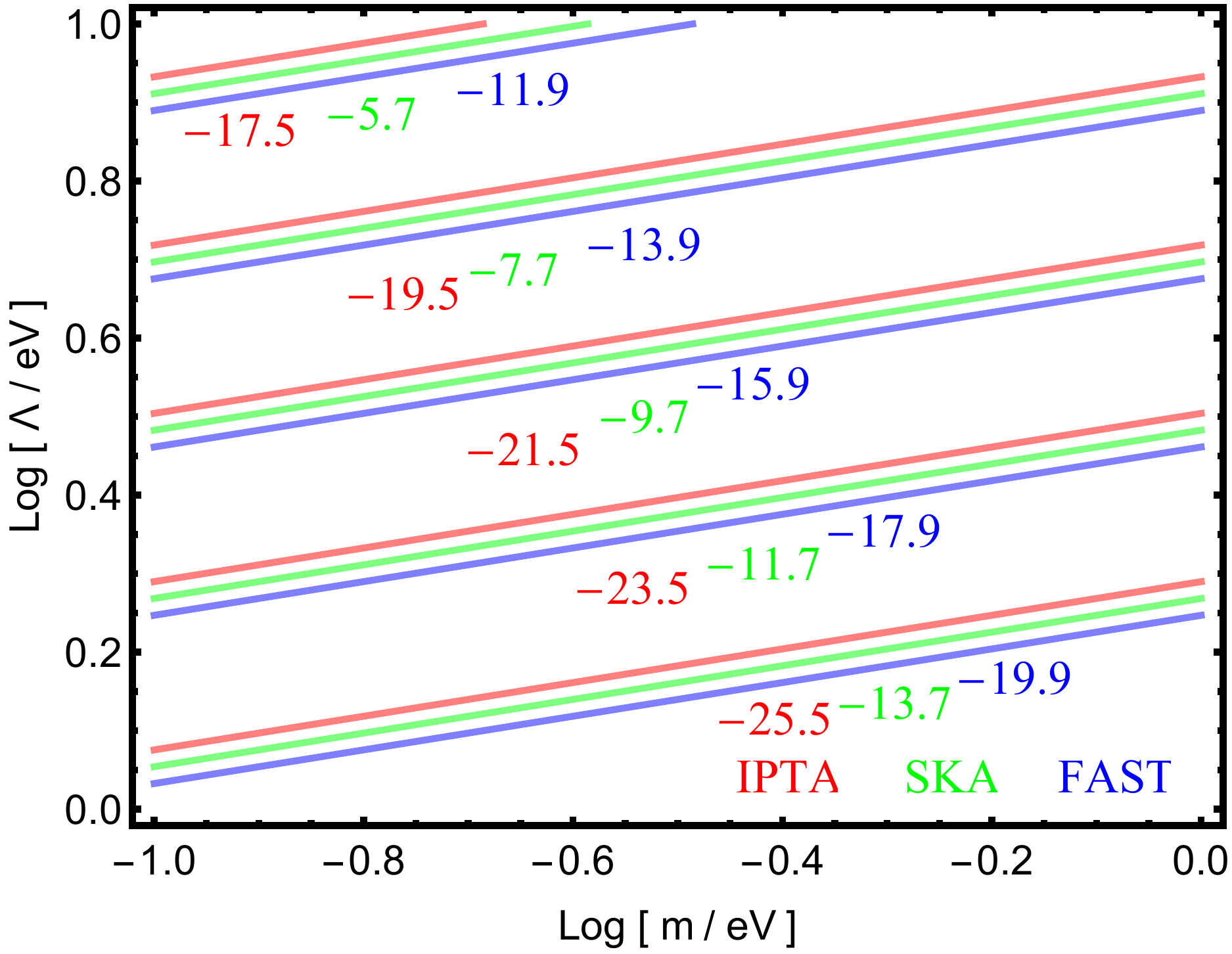}\\
  \caption{The change in the speed of GWs propagating through SfDM halo as a function of the mass of SfDM particle $m$ and the characteristic energy scale $\Lambda$ for different models in logarithmic unit. The first three lines are all for model A with superfluid phase (first line) without baryons, two-fluid phases (second line) without baryons, and superfluid phase (third line) with baryons, while the last line is for model B. Different GW detectors are labeled by various colors as clarified in the corners of each figure. The specific configurations of different GW detectors are listed in \eqref{eq:LIGO}, \eqref{eq:ET}, \eqref{eq:LISA}, \eqref{eq:BBO}, \eqref{eq:IPTA}, \eqref{eq:FAST}, \eqref{eq:SKA}, respectively.}\label{}\label{fig:dcg}
\end{figure*}

Second, what can be read from Fig.\ref{fig:dcg} ? In Fig.\ref{fig:dcg}, we present the $\delta c_g$ with respect to the SfDM model parameters $m$ and $\Lambda$ all in logarithmic unit. Different GW sources with strain $h$ and frequency $f$ are specified in the panels by the typical configurations of GW detectors. DO not interpret the numbers of contours as the detection ability of GW detectors. They are the sensitivity numbers required for the GW detectors in order to probe that part of parameter spaces of SfDM models. The first three lines present the SfDM model A in section Sec. \ref{subsec:model1}, where Sec. \ref{subsubsec:DMonly}, Sec. \ref{subsubsec:twofluid}, Sec. \ref{subsubsec:addbaryon} are presented in order in the first, second and third lines, respectively. The last line presents the SfDM model B in Sec. \ref{subsec:model2}.

The ground-based GW detectors, like LIGO and ET, have to reach a sensitivity of $10^{-60}\lesssim\delta c_g\lesssim10^{-40}$ to explore the relevant parameter spaces of SfDM model A, which is difficult even with the help of multimessenger astrophysics. The future space-borne GW detectors, like LISA and BBO, have to reach a sensitivity of $10^{-40}\lesssim\delta c_g\lesssim10^{-30}$ to explore the relevant parameter spaces of SfDM model A, which is promising with help of electromagnetic counterpart. The GW detectors that are sensitive around nHz can probe most of parameter space of interest of SfDM model A with sensitivity of $10^{-35}\lesssim\delta c_g\lesssim10^{-10}$. Likewise, the same comments for SfDM model A also apply to model B, but with much more promising perspective. However, unlike the model A, decreasing the mass of axionlike particle makes it easier for GW probe in SfDM model B.

Third, we will briefly discuss the Shapiro time delay \cite{Shapiro:1964uw} when GWs and photons encounter the gravitational potential of DM along the line of sight. Considering GWs with relatively high frequency, which are relevant for ground-based detectors, the geometrical approximation holds and we can apply the standard formula
\begin{align}\label{eq:Shapiro delay}
\Delta t=(1+\gamma_{\rm PPN})GM\ln\bigg(\frac{D}{b}\bigg).
\end{align}
where $\gamma_{\rm PPN}$ is parametrized post-Newtonian (PPN) parameter, $D$ is the distance to source and $b$ is impact parameter. In this case, both GWs and photons share the same time delay, which is given by \eqref{eq:Shapiro delay}.

It is, however, another story when the wavelengths of GWs are larger than the size of the lensing object, i.e. $\lambda_{\rm GW}\gtrsim GM/c^{2}$, which can be rewritten for the lensing mass as $M\lesssim 10^{5}M_{\odot}(f/\rm Hz)^{-1}$ \cite{Kahya:2016prx, Takahashi:2016jom}. In this case, the geometrical approximation breaks down and we have to take its wave optics into account. The additional time delay between GW and EM signals can reach $\sim 0.1 \rm s (f/\rm Hz)^{-1}$. Since the mass of DM halo we take is $M\sim 10^{12}M_{\odot}$, for GWs with frequencies relevant for LIGO, ET, LISA and BBO, the geometrical approximation remains valid and we do not have to consider the arrival time difference. As for those frequencies down to $10^{-7}\rm Hz$, the additional time delay should be considered when multimessenger analysis is involved.

\section{Conclusions} \label{sec:conclusion}

In this paper, we have studied the possibility that probing the relevant parameter space of SfDM with GWs. The results we obtained indicate that ground-based GW detectors, like LIGO and ET, are difficult to put constraints on the parameters for all models even with the help of multimessenger approach. As for space-borne GW detectors, like LISA and BBO, the ability to constrain the parameter space will be improved with the help of electromagnetic counterpart. The GW detectors sensitive around nHz, including IPTA, FAST, and SKA, are shown to be the most promising tools to probe most of parameter space of interest.

Two comments follow. First, how to distinguish the velocity changes of GWs through BEC from those due to massive graviton ? The velocity changes of GWs from massive graviton are universal independent of the GW sources and DM halos during propagation, however, one should otherwise observe different patterns of velocity changes of GWs through different DM halos from different GW sources at different sky locations. Second, a recent paper \cite{Boran:2017rdn} claims to rule out the dark matter emulators scenarios, like Bekenstein's TeVeS theory \cite{Bekenstein:2004ne} and Moffat's Scalar-Tensor-Vector gravity theory \cite{Moffat:2005si}. However, the SfDM models as MOND emulators scenarios are not ruled out yet. We hope our work will shed light on the test of SfDM scenario with GWs in future.

\begin{acknowledgments}
We thank Sebastian Ohmer for the helpful correspondence and Shantanu Desai for useful comments. This work is supported by the National Natural Science Foundation of China Grants No.11690022, No.11375247, No.11435006, and No.11647601, and by the Strategic Priority Research Program of CAS Grant No.XDB23030100 and by the Key Research Program of Frontier Sciences of CAS. This paper is dedicated for memory of the chief scientist Prof. Rendong Nan of FAST project, whose death is a great loss of Chinese Astrophysics Community.
\end{acknowledgments}

\bibliographystyle{utphys}
\bibliography{ref}

\end{document}